\documentclass[12pt]{article}
\usepackage[english]{babel}
\usepackage{graphicx,epsfig}
\makeindex \textwidth 170mm \textheight 220mm \topmargin -5mm
\newcommand{\be}{\begin{equation}}
\newcommand{\ee}{\end{equation}}
\newcommand{\bea}{\begin{eqnarray}}
\newcommand{\eea}{\end{eqnarray}}

\def\rfr#1{eq. (\ref{#1})}

\def\eqi{\begin{equation}}
\def\eqf{\end{equation}}
\def\eqia{\begin{eqnarray}}
\def\eqfa{\end{eqnarray}}
\def\btab{\begin{tabular}}
\def\etab{\end{tabular}}
\def\bar{\begin{array}}
\def\ear{\end{array}}
\def\dert#1#2{\rp{{d}{#1}}{{d}{#2}}}
\def\GR{General Relativity}
\def\grl{general relativistic}
\def\wfs{weak--field and slow--motion approximation}
\def\leti{Lense--Thirring}
\def\grc{gravitomagnetic}

\def\lg{{\rm LAGEOS}}
\def\lgg{{\rm LAGEOS} II}

\def\lb#1{\label{#1}}
\def\pc{precession}
\def\nd{node}
\def\pg{perigee}

\def\nl{nodal}

\def\sa{semimajor axis}
\def\ec{eccentricity}
\def\ic{inclination}

\def\et{Earth}
\def\ef{effect}
\def\dt#1{\dot{#1}}
\def\mlt{{\rm \mu_{LT}}}

\def\st{satellite}
\def\lt{_{\rm{LT}}}
\def\rp#1#2{{#1\over#2}}

\begin{document}
\begin{titlepage}
\begin{flushright}
\today\\
BARI-TH/00\\
\end{flushright}
\vspace{.5cm}
\begin{center}
{\LARGE On a new observable for measuring the Lense--Thirring
effect with Satellite Laser Ranging} \vspace{1.0cm}
\quad\\
{Lorenzo Iorio$^{\dag}$\\ \vspace{0.1cm}
\quad\\
{\dag}Dipartimento di Fisica dell' Universit{\`{a}} di Bari, via
Amendola 173, 70126, Bari, Italy\\ \vspace{0.2cm} } \vspace{0.2cm}
\quad\\

\vspace{0.2cm} \vspace{1.0cm}

{\bf Abstract\\}
\end{center}

{\noindent \small For a pair of twin Earth orbiting artificial
satellites placed in identical orbits with supplementary
inclinations, in addition to the sum of the residuals of the nodal
rates, already proposed for the LAGEOS--LARES mission, also the
difference of the residuals of the perigee rates could be
employed, in principle, for measuring the general relativistic
Lense--Thirring effect. Indeed, on one hand, the gravitomagnetic
secular precessions of the perigees of two supplementary
satellites in identical orbits are equal and opposite, and, on the
other, the classical secular precessions induced by the multipolar
expansion of the terrestrial gravitational field are equal, so
that their aliasing effect cancels out in the difference of the
perigees'rates. If the eccentricities of the two satellites would
be chosen to be equal, contrary to the LAGEOS--LARES project, such
cancellation would occur at a very accurate level. Among the
time--dependent perturbations, the proposed observable would allow
to cancel out the even and odd zonal gravitational tidal
perturbations and some non--gravitational perturbations. With a
proper choice of the inclination of the two satellites, the
periods of all the uncancelled time--dependent perturbations could
be made short enough to allow to fit and remove them from the
signal over observational time spans of a few years. The linear
perturbation induced by the terrestrial Yarkovski--Rubincam effect
would affect the proposed measurement at a level well below
10$^{-3}$.}

{\noindent \small }
\end{titlepage} \newpage \pagestyle{myheadings} \setcounter{page}{1}
\vspace{0.2cm} \baselineskip 14pt

\setcounter{footnote}{0}
\setlength{\baselineskip}{1.5\baselineskip}
\renewcommand{\theequation}{\mbox{$\arabic{equation}$}}
\noindent
\section{Introduction}
In its \wfs\ \GR\ predicts that, among other things,  the orbit of
a test particle freely falling in the gravitational field of a
central rotating body is affected by the so called \grc\ dragging
of the inertial frames or
\leti\ \ef. More precisely,  the longitude of the ascending \nd\ $\Omega$ and
the argument of the \pg\ $\omega$ of the orbit \cite{ste} undergo
tiny secular \pc s \cite{ciuwhe} (The original papers by Lense and
Thirring can be found in english translation in \cite{let}) \eqia
\dot\Omega \lt & = &
\frac{2GJ}{c^{2}a^{3}(1-e^{2})^{\frac{3}{2}}},\\
\dot\omega \lt & = &
-\frac{6GJ\cos{i}}{c^{2}a^{3}(1-e^{2})^{\frac{3}{2}}},\lb{perigeo}\eqfa
in which $G$ is the Newtonian gravitational constant, $J$ is the
proper angular momentum of the central body, $c$ is the speed of
light $in\ vacuum$, $a,\ e$ and $i$ are the \sa, the \ec\ and the
\ic, respectively, of the orbit of the test particle.

The first experimental check of this predicted \ef\ in the
gravitational field of the \et\ has been obtained by analyzing a
suitable combination of the laser-ranged data to the existing
passive geodetic \st s \lg\ and \lgg\ \cite{ciuetal}. The claimed
total relative accuracy of the measurement of the solve-for
parameter $\mlt$, introduced in order to account for this \grl\
\ef, is of the order of\footnote{However, some scientists propose
a different error budget \cite{riesetal}. } $2\times 10^{-1}$
\cite{ciuetal}.

In order to achieve a few percent accuracy, in \cite{ciu1} it was
proposed to launch a passive geodetic laser-ranged \st- the former
LAGEOS III, subsequently become the LARES \cite{ciu3} - with the
same orbital parameters of \lg\ apart from its inclination which
should be supplementary to that of \lg.

This orbital configuration would be able to cancel out exactly the
mismodelled part of the classical \nl\ \pc s induced by the
multipolar expansion of the terrestrial gravitational field, which
are proportional to $\cos i$ and depend on even powers of $\sin
i$, provided that the observable to be adopted is the sum of the
residuals of the \nl\ \pc s of {\rm LARES} and LAGEOS \eqi
\delta\dt\Omega^{{\rm LAGEOS}}+\delta\dt\Omega^{{\rm
LARES}}=62\mlt.\lb{lares}\eqf

Currently, the observable of the LAGEOS--LARES mission is under
revision in order to improve the obtainable accuracy
\cite{ioretal}.

The orbital parameters of \lg, \lgg\ and LARES are in Tab. 1.

\begin{table}[ht!]
\caption{Orbital parameters of \lg, \lgg\ and LARES.} \label{para}
\begin{center}
\begin{tabular}{lllll}
\noalign{\hrule height 1.5pt} Orbital parameter & \lg & \lgg &
LARES\\ \hline
$a$ (km) & 12,270 & 12,163 & 12,270\\
$e$ & 0.0045 & 0.014 & 0.04\\
$i$ (deg) & 110 & 52.65 & 70\\
\noalign{\hrule height 1.5pt}
\end{tabular}
\end{center}
\end{table}
In this paper we show that the configuration of twin satellites
placed in identical orbits with supplementary inclinations can
reveal itself more fruitful than that one could have imagined
before. Indeed, the sum of the nodes can be supplemented with a
new, independent observable given by the difference of the
perigees \cite{iordiffper}. Of course, such observable would be
more difficult to be implemented because, contrary to the nodes,
the perigee is a very sensitive orbital element which is affected
by many gravitational and non--gravitational perturbations which
should be very carefully modelled and treated in the orbital
processors like GEODYN II or UTOPIA. However, the great experience
obtained in dealing with the perigee of LAGEOS II in the
LAGEOS--LAGEOS II Lense--Thirring experiment could be fully
exploited in such measurement as well.

The paper is organized as follows. In section 2 we describe such
new observable, the impact of the mismodelled static part of the
gravitational field of the Earth on it and some possibilities for
its practical implementation. In section 3 and 4 we sketch the
impact of the non--gravitational and gravitational orbital
perturbations on the proposed measurement (a more quantitative
analysis with numerical tests can be found in \cite{lkior}).
Section 5 is devoted to the conclusions.
\section{A new perigee--only observable}
The concept of a couple of satellites placed in identical orbits
with supplementary inclinations could be fruitfully exploited in
the following new way.

An inspection of \rfr{perigeo} and of the explicit expressions of
the rates of the classical perigee precessions induced by the even
zonal harmonics of the geopotential \cite{ior2} suggests to adopt
as observable the difference of the residuals of the perigee
precessions of the two satellites
\eqi\delta\dot\omega^{i}-\delta\dot\omega^{\pi-i}=X_{\rm LT
}\mu_{\rm LT},\lb{new}\eqf so to obtain a secular trend with a
slope of $X_{\rm LT}$ mas yr$^{-1}$. Indeed, on one hand, the
Lense--Thirring perigee precessions depend on $\cos i$, contrary
to the nodal rates which are independent of the inclination, so
that, by considering the relativistic effect as an unmodelled
force entirely adsorbed in the residuals, in \rfr{new} they sum
up. On the other, it turns out that the classical even zonal
perigee precessions depend on even powers of $\sin i$ and on
$\cos^2 i$, so that they cancel out exactly in \rfr{new}. It may
be interesting to notice that the proposed observable of \rfr{new}
is insensitive to the other general relativistic feature which
affects the pericenter of a test body, i.e. the gravitoelectric
Einstein precession. Indeed, as it is well known \cite{ciuwhe}, it
does not depend on the inclination of the satellite's orbital
plane.

In regard to a practical application of such idea, we note that
the LAGEOS--LARES mission would be unsuitable because the perigee
of LAGEOS is not a good observable due to the notable smallness of
the eccentricity of its orbit. For the sake of concreteness, we
could think about a LARES II which should be the supplementary
companion of LAGEOS II. In this case we would have a
gravitomagnetic trend with a slope of -115.2 mas yr$^{-1}$ (which
is almost twice that of the LAGEOS--LARES node--only mission).
Moreover, since the magnitude of the eccentricity of LAGEOS II is
satisfactory in order to perform relativistic measurements with
its perigee, the LARES II, contrary to the LAGEOS--LARES mission,
could be inserted in an orbit with the same eccentricity of that
of LAGEOS II. So, the cancellation of the classical secular
precessions would occur at a higher level than in the
LAGEOS--LARES node--only observable \cite{ioretal}. Of course, a
careful analysis of the time--dependent gravitational \cite{ior1}
and, especially, non--gravitational perturbations (see \cite{luc1}
for the radiative perturbations and \cite{luc2} for the thermal,
spin--dependent perturbations), to which the perigee is
particularly sensitive, contrary to the node, is needed in order
to make clear if also for such perturbations some useful
cancellations may occur, and to which extent the uncancelled
perturbations may affect the proposed measurement. In the
following section we will perform a preliminary investigation: a
more detailed analysis has been performed in \cite{lkior}.
\section{The non--gravitational perturbations}
\subsection{The radiative perturbations}
\subsubsection{The direct solar radiation pressure} According to
\cite{luc1}, the direct solar radiation pressure does not induce
any secular trend on the perigee rate: its signature is
long--periodic. Its effect on the difference of the perigee rates
of two supplementary satellites amounts to\footnote{In deriving
\rfr{srp} it has been accounted for the fact that for a couple of
supplementary satellites the classical rate of the node changes
sign because it depends on $\cos i$, while the rate of perigee
remains unchanged because it depends on $\cos^2 i$ and on even
powers of $\sin i$ \cite{ior2}. }
\begin{equation}
\dot\omega^{i}_{\rm SRP}-\dot\omega^{\pi-i}_{\rm
SRP}={3A_{\odot}\over {8nae}}\left\{
\begin{array}{lllll} -2\cos i\cos \epsilon
\cos \Big(\Omega+\lambda+\omega \Big) -\\
2\Big [\Big(1-\cos i\Big)\cos \epsilon\Big ]
\cos \Big(\Omega+\lambda-\omega \Big) + \\
\Big [2\Big(1+\cos i\Big)\cos \epsilon\Big ]
\cos \Big(\Omega-\lambda+\omega \Big) + \\
-2\cos i\cos \epsilon
\cos \Big(\Omega-\lambda-\omega \Big),\\
\end{array} \right.
\label{srp}
\end{equation}
In it $A_{\odot}$ is the acceleration induced by the direct solar
radiation pressure. In the case of a supplementary configuration
based on LAGEOS II the harmonic $\cos(\Omega+\lambda-\omega)$
would induce serious troubles for the proposed measurement of the
Lense--Thirring effect. Indeed, on one hand its period amounts to
4,244 days, i.e. 11.6 years, on the other, even by assuming a
0.5$\%$ mismodelling in $A_{\odot}$ \cite{luc1}, the mismodelled
amplitude of the perigee rate amounts to 609 mas yr$^{-1}$, while
the Lense--Thirring effect is, for the LAGEOS II supplementary
configuration, 115.2 mas yr$^{-1}$. The mismodelled amplitude of
the perigee perturbation would amount to 750.8 mas. Then, over a
reasonable time span of a few years it would superimpose to the
relativistic signal and its level of uncertainty would vanish any
attempts for extracting the gravitomagnetic signature.

The situation ameliorates if we consider a couple of entirely new
laser--ranged satellites of LAGEOS--type with frozen perigees.
This means that the inclination would amount to $63.4^{\circ}$, so
to make the period of perigee extremely long.
\begin{table}[ht!]
\caption{Orbital parameters of the new supplementary satellites.}
\label{para}
\begin{center}
\begin{tabular}{ccccccc}
\noalign{\hrule height 1.5pt} $a$ (km) & $i$ (deg) & $e$ &
$P(\Omega)$ (days) & $P(\omega)$ (days) & $\dot\Omega_{\rm LT}$
(mas/y) &  $\dot\omega_{\rm LT}$ (mas/y)\\
\hline
12,000 & 63.4 (116.6)   & 0.05 & $\mp$733.45  & 269,078.41 & 33 & $\mp$44.4\\
\noalign{\hrule height 1.5pt}
\end{tabular}
\end{center}
\end{table}
A possible orbital configuration could be that in Table 2. In it
we quote also the Lense--Thirring effects on the node and the
perigee. So, the periodicities of the perturbing harmonics would
amount to
\begin{eqnarray}
P(\Omega+\lambda-\omega)&=&P(\Omega+\lambda+\omega)=729.56\ {\rm
days},\\
P(\Omega-\lambda-\omega)&=&P(\Omega-\lambda+\omega)=-243.8\ {\rm
days}.
\end{eqnarray}
This is very important because, in this case, over an
observational time span of a few years the time--dependent
perturbations due to the direct solar radiation pressure could be
viewed as empirically fitted quantities and could be removed from
the signal.

It should be noticed that the practical data reduction of the
perigee rates should be performed very carefully in order to
account for possible, unpredictable changes in the physical
properties of the satellites' surfaces which may occur after some
years of their orbital life, as it seems it has happened for
LAGEOS II. Such effects may yield a not negligible impact on the
response to the direct solar radiation pressure. However, the
great experience obtained in dealing with the perigee of LAGEOS II
in the LAGEOS--LAGEOS II Lense--Thirring experiment could be fully
exploited  for the proposed measurement as well.
\subsubsection{The Earth's albedo} For the Earth's albedo, which
induces only long--periodic harmonic perturbations on the perigee
rate \cite{luc1}, the same considerations as for the direct solar
radiation pressure hold because the periodicities of the harmonic
constituents are the same.
\subsection{The thermal perturbations}
\subsubsection{The Yarkovski--Rubincam effect} According to \cite{luc2}, the terrestrial Yarkovski--Rubincam effect
induces on the perigee rate both secular and long--periodic
perturbations. Of course, in regard to the measurement of the
secular Lense--Thirring trend the linear Rubincam effect is the
most insidious one.

The genuine secular part of the terrestrial Yarkovski--Rubincam
perturbation on the perigee rate is, according to \cite{luc2} \eqi
\dot\omega_{\rm Rub\ sec}=\rp{A_{\rm
Rub}}{4na}\cos\vartheta\left[1+2\cos^2 i+S_z^2 \left(1-6\cos^2
i\right)\right],\eqf where $A_{\rm Rub}$ is the acceleration due
to the Rubincam effect, $\vartheta$ is the thermal lag angle and
$S_z$ is the component of the satellite's spin axis along the $z$
axis of a geocentric, equatorial inertial frame.

By assuming, for the sake of generality, for the supplementary
satellite a thermal lag angle slightly different from that of its
twin, so that $\vartheta^{\pi-i}=\vartheta^{i}+\delta$ with
$\delta$ small, the Rubincam secular effect on the difference of
the perigee rates of a pair of supplementary satellites becomes
\begin{eqnarray}\dot\omega^{i}_{\rm Rub\ sec}-\dot\omega^{\pi-i}_{\rm
Rub\ sec}&=&\rp{A_{\rm Rub}}{4na}\left\{\left[\left(1+2\cos^2
i\right)+\left(1-6\cos^2 i \right)\left(S_z^{\pi-i}\right)^2\right]\delta\sin\vartheta +\right.\nonumber\\
&+& \left.\left(1-6\cos^2
i\right)\cos\vartheta\left[\left(S_z^{i}\right)^2-\left(S_z^{\pi-i}\right)^2\right]\right\}.\lb{rub1}\end{eqnarray}
By assuming $\left(S_z^{i}\right)^2=\left(S_z^{\pi-i}\right)^2$
and $\left(S_z^{\pi-i}\right)^2 =1$, \rfr{rub1} reduces to
\eqi\dot\omega^{i}_{\rm Rub\ sec}-\dot\omega^{\pi-i}_{\rm Rub\
sec}=\rp{A_{\rm Rub}}{4na}\left(2-4\cos^2
i\right)\delta\sin\vartheta.\lb{rub2}\eqf According to
\cite{luc2}, for LAGEOS II $A_{\rm Rub}=-6.62\times 10^{-10}$ cm
s$^{-2}$, so that \eqi\rp{A_{\rm Rub}}{4na}=-1.8\ {\rm mas\
yr^{-1}}.\eqf By assuming $\vartheta= 55^{\circ}$ , as for LAGEOS,
and a mismodelling of 20$\%$ on $A_{\rm Rub}$, \rfr{rub2}, for an
orbital supplementary configuration based on LAGEOS II, yields a
mismodelled linear trend of $\delta\times 0.1$ mas yr$^{-1}$, so
that the relative error in the Lense--Thirring measurement would
amount to $\delta\times (1.4\times 10^{-3})$.

On the other hand, by assuming $\delta\sim 0^{\circ}$  and
$\left(S_z^{i}\right)^2\neq\left(S_z^{\pi-i}\right)^2$, \rfr{rub1}
reduces to \eqi\dot\omega^{i}_{\rm Rub\
sec}-\dot\omega^{\pi-i}_{\rm Rub\ sec}=\rp{A_{\rm
Rub}}{4na}\left(1-6\cos^2
i\right)\cos\vartheta\left[\left(S_z^{i}\right)^2-\left(S_z^{\pi-i}\right)^2\right].\lb{rub3}\eqf
With the same assumption as before for LAGEOS II, \rfr{rub3}
yields $\left[\left(S_z^{\rm L2}\right)^2-\left(S_z^{\rm
LR2}\right)^2 \right]\times 0.2$ mas yr$^{-1}$ with a relative
error in the Lense--Thirring measurement of $\left[\left(S_z^{\rm
L2}\right)^2-\left(S_z^{\rm LR2}\right)^2 \right]\times (1.7\times
10^{-3})$. However, it should be noticed that both $\delta$ and
$\left(S_z^{i}\right)^2-\left(S_z^{\pi-i}\right)^2$ could be made
very small, so that the presented estimates would become even more
favorable. If, e.g. we think about a pair of new, geodetic
satellites of LAGEOS type constructed very carefully in the same
way and placed in supplementary frozen perigee orbital
configuration, it would be quite reasonable to assume their spins
as directed mainly along the $z$ axis, during the first years of
orbital life, as it happened for LAGEOS and LAGEOS II. Numerical
simulations confirm such feature. However, it is also important to
notice that the difference of the squares of $S_z$ is involved, so
that possible inversions of the rotational motion of the
satellites would not yield problems.

According to \cite{luc2}, among the periodicities of the harmonic
terms $2\dot\omega$ and $4\dot\omega$ are present. For a couple of
new supplementary satellites with frozen perigees the mismodelled
part of such harmonics, which, in general do not cancel out in the
difference of the perigee rates, would resemble aliasing secular
trends. However, their impact on the Lense--Thirring measurement
should be at the 10$^{-3}$ level because their amplitude is
proportional to $\rp{(\delta A_{\rm Rub})}{4na}$ which amounts to
0.3 mas yr$^{-1}$, for LAGEOS II, by assuming a mismodelling of
20$\%$ in $A_{\rm Rub}$.
\subsubsection{The solar Yarkovski--Schach effect} The solar
Yarkovski--Schach effect does not induce secular perturbations on
the perigee rate \cite{luc2}. In regard to its long--periodic
harmonic terms, which, in general, do not cancel out in the
difference of the rates of the perigees of a pair of supplementary
satellites, if the frozen perigee configuration would be adopted,
their impact was not insidious for the proposed Lense--Thirring
measurement. Indeed, as can be inferred from Table 4 of
\cite{luc2}, their periodicities do not contain any multiple of
the perigee frequency, so that, with the proposed configuration,
no semisecular terms would affect the signal.
\section{The gravitational tidal perturbations}
In regard to the orbital perturbations induced by the Earth solid
and ocean tides, according to \cite{ior1}, the perigee is
particularly sensitive to them, not only to the $l=2$ part of the
tidal spectrum, but also to the $l=3$ constituents.

An important role in assessing the impact of the long periodic
tidal perturbations on the proposed measurement of the
Lense--Thirring effect is played by their frequencies which are
given by
\eqi\dot\Gamma_f+(l-2p)\dot\omega+m\dot\Omega.\lb{fqz}\eqf Recall
that \cite{ior1} for the even constituents $l-2p=0$, while for the
odd constituents $l-2p\neq 0$. Moreover, in \rfr{fqz}, for a given
tidal constituent of frequency $f$, $\dot\Gamma_f$ depends only on
the luni--solar variables. In order to evaluate correctly the
impact of the gravitational time--dependent perturbations on the
perigee, it is important to note that, according to eq. (36) and
eq. (50) of \cite{ior1}, their amplitudes are proportional to
\eqi\rp{(1-e^2)}{e}F_{lmp}\dert{G_{lpq}}{e}-\rp{\cos i}{\sin
i}G_{lpq}\dert{F_{lmp}}{i},\eqf where $F_{lmp}(i)$ and
$G_{lpq}(e)$ are the inclination functions and eccentricity
functions, respectively \cite{kau}.

Fortunately, the proposed combination
$\dot\omega^{i}-\dot\omega^{\pi-i}$ allows to cancel out the
18.6--year and the 9.3--year tides because they are even zonal
perturbations. This is an important feature because their
extremely long  periods are independent of those of the node
and/or the perigee of the satellites to be employed: indeed, they
depend only on the luni--solar variables. Moreover, their $l=2,\
m=0$ constituents would have large amplitudes, so that, if not
canceled out, they would represent very insidious superimposed
biasing trends. In regard to the solid and ocean\footnote{For the
ocean tides we consider only the prograde constituents.} $l=2$
tesseral ($m=1$) and sectorial ($m=2$) tides, from the fact that
their frequencies depend on $\dot\Omega$, which, as already
previously pointed out, changes sign for a supplementary
satellite, and from
\begin{eqnarray}
F_{211}&=&-\rp{3}{2}\sin i \cos i,\\
\dert{F_{211}}{i} &=&-\rp{3}{2}\left(\cos^2 i-\sin^2 i\right),
\\
F_{221}&=&\rp{3}{2}\sin^2 i,\\
\dert{F_{221}}{i} &=&3\sin i\cos i,
\\
\end{eqnarray} it turns out that they do affect
$\dot\omega^i -\dot\omega^{\pi-i}$. However, this fact would not
have a serious impact on the proposed measurement of the
Lense--Thirring effect since the periods of such perturbations
would not be too long, so that they could be fitted and removed
from the signal over an observational time span of some years.

A careful analysis must be performed for the ocean odd tidal
perturbations.  Fortunately, the odd zonal ($l=3,\ m=0$) tidal
perturbations cancel out. Indeed, for them it turns out that
\begin{eqnarray}
F_{300}&=&-\rp{5}{16}\sin^3 i,\\
\dert{F_{300}}{i}&=&-\rp{15}{16}\sin^2 i\cos i,\\
F_{301}&=&\rp{15}{16}\sin^3 i-\rp{3}{4}\sin i,\\
\dert{F_{301}}{i}&=&\rp{45}{16}\sin^2 i\cos i-\rp{3}{4}\cos i,\\
F_{302}&=&-F_{301},\\
\dert{F_{302}}{i}&=&-\dert{F_{301}}{i},\\
F_{303}&=&-F_{300},\\
\dert{F_{303}}{i}&=&-\dert{F_{300}}{i}.
\end{eqnarray}
Moreover, their frequencies are independent of $\dot\Omega$.

It is important to notice that the same result holds also for the
time--dependent perturbation induced on the perigee by the
$J_{2n+1}$ odd zonal harmonics of the geopotential; the $J_3$
constituent, e.g., depends on odd powers of $\sin i$, as it can be
noted by the explicit expression of eq. (18) of \cite{stokaz}.
This is an important feature because their frequencies are
multiple of $\dot\omega$; for a frozen perigee configuration they
might represent very insidious secular perturbations.

In regard to the $l=3$ tesseral ($m=1$) and sectorial ($m=2$)
tidal lines, it can be proved that they do affect
$\dot\omega^{i}-\dot\omega^{\pi-i}$ because their amplitudes
depend on inclination functions $F_{lmp}(i)$ which depend, among
other factors, also on $\cos i$, contrary to the $l=3,\ m=0$ case.
Indeed, it turns out
\begin{eqnarray}
F_{311}&=&\rp{15}{16}\sin^2 i (1+3\cos i)-\rp{3}{4}(1+\cos i),\\
\dert{F_{311}}{i}&=&\rp{15}{8}\sin i\cos i(1+3\cos i)-\rp{45}{16}\sin^3 i+\rp{3}{4}\sin i,\\
F_{312}&=&\rp{15}{16}\sin^2 i (1-3\cos i)-\rp{3}{4}(1-\cos i),\\
\dert{F_{312}}{i}&=&\rp{15}{8}\sin i\cos i(1-3\cos i)+\rp{45}{16}\sin^3 i-\rp{3}{4}\sin i,\\
F_{321}&=&\rp{15}{8}\sin i(1-2\cos i-3\cos^2 i),\\
\dert{F_{321}}{i}&=&\rp{15}{8}\cos i(1-2\cos i-3\cos^2 i)+\rp{15}{8}\sin i(2\sin i+6\sin i\cos i),\\
F_{322}&=&-\rp{15}{8}\sin i(1+2\cos i-3\cos^2 i),\\
\dert{F_{322}}{i}&=&-\rp{15}{8}\cos i(1+2\cos i-3\cos^2
i)-\rp{15}{8}\sin i(-2\sin i+6\sin i\cos i).
\end{eqnarray}
Moreover, their frequencies are combinations of the form
$\dot\Gamma_f\pm\dot\omega+\dot\Omega$ and
$\dot\Gamma_f\pm\dot\omega+2\dot\Omega$. For an orbital
configuration based on LAGEOS II this fact would represent a
serious drawback because, as pointed out in \cite{ior1}, the
$K_1,\ l=3,\ m=1,\ p=1,\ q=-1$ tidal line induces on the perigee
of LAGEOS II  a perturbation with nominal amplitude of -1,136 mas
and period of 1,851.9 days, i.e. 5.07 years, from the frequency
$\dot\omega+\dot\Omega$. The situation is quite similar to that of
the direct solar radiation pressure harmonic with a period of 11.6
years. Instead, as in that case, with a frozen perigee
configuration the period of the $K_1,\ l=3,\ m=1,\ p=1,\ q=-1$
tidal line would greatly reduce, so that it could be fitted and
removed from the signal over an observational time span of a few
years. The same holds also for the other $l=3,\ m=1,2$ tidal
lines.
\section{Conclusions}
In this paper we have proposed to consider the difference of the
residuals of the perigee rates $\delta\dot\omega^i
-\delta\dot\omega^{\pi-i}$, in addition to the already proposed
sum of the residuals of the nodes $\delta\dot\Omega^i
-\delta\dot\Omega^{\pi-i}$, as  a new observable for measuring the
Lense--Thirring effect with a pair of laser--ranged Earth's
satellites of LAGEOS--type in identical orbits with supplementary
inclinations. In the well known originally proposed LAGEOS --LARES
mission $\delta\dot\omega^i -\delta\dot\omega^{\pi-i}$ would be
unsuitable because the perigee of LAGEOS cannot be measured
accurately due to the smallness of its eccentricity. It should be
pointed out that in this paper we have not intended to propose the
launch of a couple of new LAGEOS--like satellites in order to
measure the difference of the Lense--Thirring perigees'rates
instead of considering the sum of the nodes of the LAGEOS--LARES
mission. We have simply investigated, in a preliminary way and
from a scientific point of view, if it would have sense to
consider the difference of the perigees as a possible,
complementary and independent observable for measuring the
Lense--Thirring effect with respect to the sum of the nodes: of
course, with a couple of new SLR satellites we would have at our
disposal both the observables. As it has been shown more
quantitatively in \cite{lkior}, the chosen orbital configuration
for the two satellites would yield great benefits also to the
measurement of the sum of the Lense--Thirring nodal rates which,
of course, would remain much more accurately measured.

A preliminary analysis of the systematic errors induced by the
non--gravitational and gravitational orbital perturbations has
been carried out. In regard to the gravitational perturbations,
also for $\delta\dot\omega^i -\delta\dot\omega^{\pi-i}$, as for
$\delta\dot\Omega^i +\delta\dot\Omega^{\pi-i}$, the main
systematic error induced by the mismodelled even zonal
coefficients of the multipolar expansion of the Earth's
gravitational field cancels out. Also the time--dependent odd
zonal harmonics of the geopotential would be cancelled out by the
proposed combination. Moreover, the even and odd zonal
time--varying tidal perturbations do not affect the proposed
observable. This is a very important feature because among them
there are the very insidious semisecular 18.6--year and 9.3--year
tides, whose frequencies are independent of the satellite's
orbital configuration because they depend only on the luni--solar
variables. On the contrary, the tesseral and sectorial tides and
most of the non--gravitational time--dependent perturbations do
affect $\delta\dot\omega^i -\delta\dot\omega^{\pi-i}$. It is also
very important to point out that in the near future the new
terrestrial gravity models from CHAMP and GRACE missions will be
available, so that the role of the gravitational perturbations in
the error budget will notably reduce.

If an orbital configuration based on LAGEOS II and a twin of its,
say LARES II, was adopted it would present some drawbacks because
of two uncancelled long--periodic harmonic perturbations which
have periods of 5.07 years ($K_1,\ l=3,\ m=1,\ p=1,\ q=-1$ tide)
and 11.6 years (direct solar radiation pressure), respectively.
Indeed, over observational time spans of a few years they would
resemble aliasing superimposed trends which could bias the
recovery of the linear Lense--Thirring signal. The optimal choice
would be the use of a couple of entirely new geodetic
laser--ranged satellites of LAGEOS type accurately constructed in
an identical manner with a small area--to--mass ratio, so to
minimize the impact of the non--gravitational perturbations, and
placed in a frozen perigee configuration. In this way there would
not be semi--secular effects and all the time--dependent
perturbations affecting the proposed observable would have short
enough periods so to be fitted and removed from the signal over
reasonable time spans.

The terrestrial Yarkovski--Rubincam effect would induce, among
other things, an uncancelled, genuine linear perturbation. This
fact is very important because it could mimic the relativistic
trend and make its measurement impossible due to the related level
of mismodelling. However, its impact on the proposed measurement
of the Lense--Thirring effect would be well below the 10$^{-3}$
level.

However, it should also be considered that the evaluation of the
impact of the non--gravitational perturbations on the proposed
observable has been worked out on the basis of the currently known
status of the physical properties and of the orbital geometries of
the existing LAGEOS and LAGEOS II satellites. With a couple of new
SLR satellites suitably built up, of course, it would be possible
to further reduce the non--gravitational perturbations acting on
them for example by reducing their area--to--mass ratio and using
rather eccentric orbits. Moreover, it would also possible to use
the data collected just in the first years of their lifetimes so
that certain simplifying assumptions on their spin motions could
be safely done. The great experience maturated with the
LAGEOS--LAGEOS II Lense--Thirring experiment in dealing with the
time--varying reflectivity properties of LAGEOS II and their
impact on the perigee evolution could be fully exploited as well.
According to the extensive and conservative numerical analysis of
\cite{lkior}, the total impact of the non-gravitational
perturbations should amount to almost 5$\%$. Last but not least,
the concept of supplementary satellites in identical orbits could
be implemented also with a pair of drag--free satellites thanks to
recent developments of such technique which could assure a
lifetime of many years (M.C.W. Sandford, private communication,
2002).

In conclusion, the proposal of measuring the Lense--Thirring
effect with a supplementary pair of satellites turns out to be
enforced, at least in principle, because it would be possible to
analyze not only $\delta\dot\Omega^i +\delta\dot\Omega^{\pi-i}$,
as in the originally proposed LAGEOS--LARES proposal, but also
$\delta\dot\omega^i -\delta\dot\omega^{\pi-i}$, provided that a
carefully selected orbital configuration is adopted. Moreover, if
the new satellites to be launched had rather eccentric orbits, we
would have at our disposal both their perigees, apart from that of
the existing LAGEOS II, in order to built up suitable combinations
of orbital residuals, including also the nodes, which would allow
to cancel out many mismodelled even zonal harmonic coefficients,
as done in the LAGEOS--LAGEOS II Lense--Thirring experiment.
Finally, we would also be able to perform other gravitational
tests concerning, e.g., the realtivistic gravitoelectric perigee
advance \cite{iorppn} and the hypothesis of a fifth force
\cite{ioryuk}.
\section*{Acknowledgements}
I'm grateful to L. Guerriero for his support while at Bari.
Special thanks to D.M. Lucchesi for his helpful and important
informations on the non--gravitational perturbations on LAGEOS II.

\end{document}